\documentclass[sigplan,screen, nonacm=true]{acmart}
\usepackage{algorithm}
\usepackage{algpseudocode}
\usepackage{balance}
\usepackage{booktabs}
\usepackage{bookmark}
\usepackage{comment}
\usepackage{datetime}
\usepackage{enumerate}
\usepackage{fancyhdr}
\usepackage{framed}
\usepackage{graphicx}
\usepackage{hyperref}
\usepackage[procnames]{listings}
\usepackage{microtype}
\usepackage{multirow}
\usepackage{subcaption}
\usepackage[nounderscore]{syntax}
\usepackage{textgreek}
\usepackage{enumitem}
\usepackage[normalem]{ulem}
\usepackage{url}
\usepackage{xcolor}
\usepackage{xspace}

\newcommand{\ignore}[1]{}

\graphicspath{{figures/}}
\DeclareGraphicsExtensions{.pdf,jpeg,.png}

\lstdefinestyle{mystyle}{
    basicstyle=\ttfamily\small,
    breakatwhitespace=false,         
    breaklines=true,                 
    captionpos=b,                    
    keepspaces=false,                 
    numbersep=2pt,                  
    showspaces=false,                
    showstringspaces=false,
    showtabs=false,                  
    tabsize=2
}
\lstdefinelanguage{rosette}{
  morekeywords=[1]{verify,solve,forall,and,or,assert,s-exp,rosette,set!,begin,define,define-values,define-syntax,define-syntax-rule,syntax-rules,let,let*,if,when,unless,match-define,lambda,provide,cond,case,else,struct,letrec,for/list,true,false,null,local,require,rename-in,??,define-symbolic,define-symbolic*,not,=>,ite,\#lang,\#:transparent,\#:mutable,equal?,match,list,for},
  morekeywords=[2]{bv,bvult,bvuge,bvsub,~>,bitvector,integer?},
  morekeywords=[3]{serval:split-pc,serval:bug-on},
  alsoletter={\#,:,?,-,=>},
  morecomment=[l]{;},
}

\setcopyright{none}

\begin{document}

\title[SCHE]{Homomorphically Encrypted Computation using Stochastic Encodings}

\author{Hsuan Hsiao$^\dagger$, Vincent Lee$^\dagger$, Brandon Reagen$^{\dagger\ddagger}$, Armin Alaghi$^\dagger$}
\affiliation{
\institution{$^\dagger$Reality Labs Research, $^\ddagger$New York University}
}
\email{{jhsiao, vtlee, reagen, alaghi}@fb.com,  bjr5@nyu.edu}

\begin{abstract}

Homomorphic encryption (HE) is a privacy-preserving technique that enables computation directly over ciphertext.
Unfortunately, a key challenge for HE is that implementations can be impractically slow and have limits on computation that can be efficiently implemented.
For instance, in Boolean constructions of HE like TFHE, arithmetic operations need to be decomposed into constituent elementary logic gates to implement so performance depends on logical circuit depth.
For even heavily quantized fixed-point arithmetic operations, these HE circuit implementations can be slow.

This paper explores the merit of using stochastic computing (SC) encodings to reduce the logical depth required for HE computation to enable more efficient implementations.
Contrary to computation in the plaintext space where many efficient hardware implementations are available, HE provides support for only a limited number of primitive operators and their performance may not directly correlate to their plaintext performance.
Our results show that by layering SC encodings on top of TFHE, we observe similar challenges and limitations that SC faces in the plaintext space.
Additional breakthroughs would require more support from the HE libraries to make SC with HE a viable solution.

\end{abstract}

\maketitle

\section{Introduction}
\label{sec:introduction}

\textbf{Homomorphic encryption} (HE) is a rapidly emerging privacy-preserving technology that enables computation directly over ciphertext.
This means that a potentially untrusted third party can process data without ever observing the plaintext contents.
Unfortunately, one of the key challenges with HE is that it is orders of magnitude slower than plaintext computation. 
This is because in HE any target computation needs to be decomposed into elementary gates or a limited set of arithmetic operations.
Each operation in HE is also orders of magnitude slower than the plaintext version; to close the gap, HE requires optimizations from across the computing stack to bring down to practical speeds.

\textbf{Stochastic computing} (SC)~\cite{gaines69, poppelbaum} is a computing paradigm that allows for simple and efficient implementations of basic arithmetic operations such as multiplication and addition.
Values are encoded as probabilistic bitstreams of unary values as opposed to wide fixed-point or floating-point representations.
An $n$-bit binary number $x$ would be encoded as a bitstream $X$ of length $N=2^n$ to achieve the same precision where $P(X = 1) = x/2^n$ at each position in the bitstream.
Stochastic encodings (SE) typically require an overall higher number of bits to represent a similar precision compared to conventional binary encodings (BE); 
however, the low circuit complexity (i.e., size and logical depth) of its operators and the predominantly bitwise computation style provides flexibility and high parallelization potential.

\textbf{Boolean HE.}
This work focuses on the Boolean constructions of HE such as the TFHE variant~\cite{tfhe} where bits of the data are encrypted individually and computed at the bit level.
The performance of the HE computation then scales with the complexity of the equivalent circuit required to implement operations.
For instance, to execute a 4-bit addition in TFHE, we would decompose it into the elementary AND, OR, and NOT logic gates to construct the circuit and execute each of these in TFHE.
Each gate must be executed in HE so the size and logical depth of the equivalent circuit directly impacts the run time performance of the HE computation.

\begin{figure*}
\centering
\includegraphics[width=0.8\linewidth]{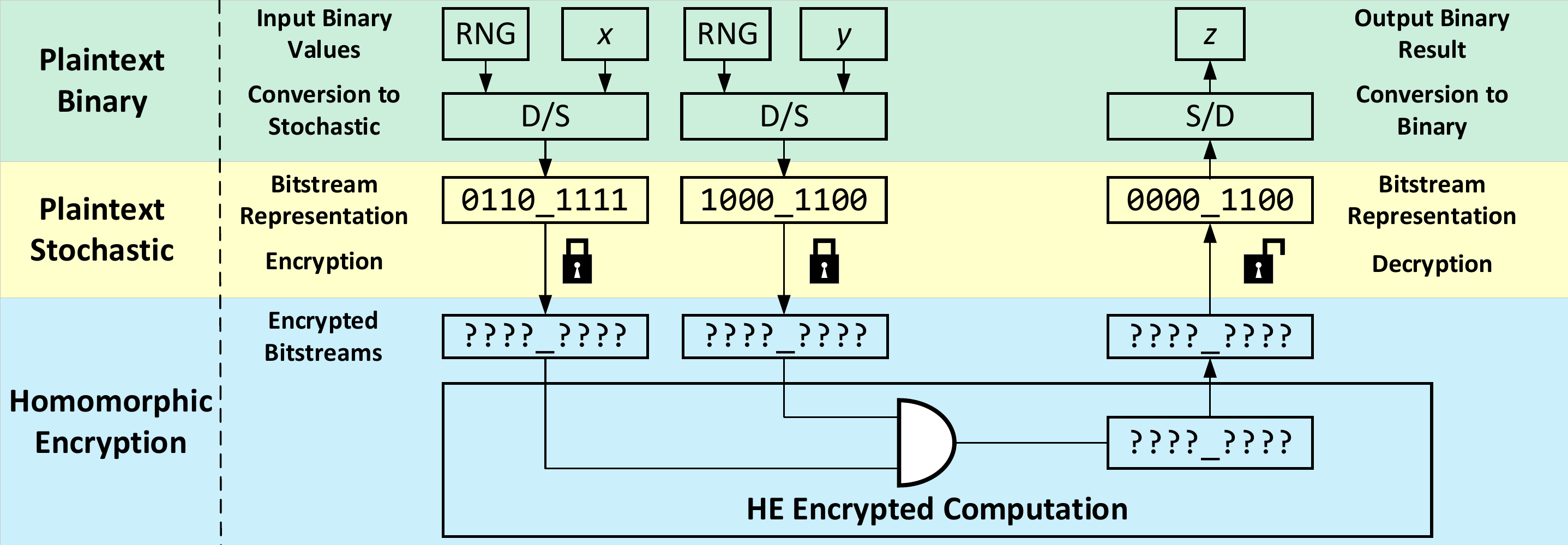}
\caption{Our proposed SCHE system which layers SE on top of HE to reduce the complexity of HE computation.  
Plaintext values are compared against a random number generator (RNG) by a digital-to-stochastic (D/S) converter. 
The resulting bitstream is then encrypted to HE and processed. 
By encoding values in SE, the size and logical depth of HE computation is reduced. 
The resulting encrypted bitstream is decrypted and converted to a digital value using a stochastic-to-digital (S/D) converter.
}
\label{fig:system}
\end{figure*}

\textbf{Proposed SCHE.}
We propose applying stochastic encodings and computing prior to encrypting data to reduce the complexity of the circuit that must be implemented in HE, which in turn reduces the computational cost.
The high level system definition and how data and computation are processed are shown in \autoref{fig:system}.

There are a few reasons why this SCHE system may be desirable.
The logical depth of arithmetic circuits for binary-encoded data typically scales with the input precision (i.e., $O(n)$), whereas SC circuits typically have a constant logical depth (i.e., $O(1)$).
Intuitively, applying SE to HE should simplify the computation and reduce overall latency.
Additionally, SC requires more bits to pass through the same circuit compared to BE computation, which requires fewer bits to pass through a more diverse circuit.
This characteristic, along with SC's potential for high parallelism, allows SC to take better advantage of executing in a Single Instruction Multiple Data (SIMD) fashion than BE computation.
Given the benefits brought about by SE and the fact that HE computation brings different cost metrics as opposed to plaintext computation, accelerating HE computation with a switch in data encoding could be potentially promising.
Additionally, any new insights we can gather about SE and SC operating in different underlying conditions than traditional hardware assumptions can be valuable to the research community.

\section{Methodology}
\label{sec:methodology}

We built our SCHE system on top of the Palisade library implementation of TFHE~\cite{PALISADE2021Lib} using 128-bit for the security parameter; our experiments are run on a 2.3GHz 8-Core Intel Core i9 processor with 32GB of 2667MHz DDR4 memory.
We evaluate and compare the performance difference between SC and conventional BE arithmetic by implementing both styles of circuits for multiply and add, which are the two main operations in HE.

Since the performance difference between using SE and BE is impacted directly by the gate-level implementation of the computation, we evaluate the upper bound on potential savings by selecting naive implementations for BE and the most performant implementations for SE.
We implement BE multiplication as an array multiplier, which consists of $6n^2$ gates and has a logical depth of $8n$ gates for $n$-bit inputs. 
For SC multiplication, we assume an AND-gate multiplier, which computes the product by bitwise AND-ing two bitstreams and has a logical depth of 1 gate.
For addition in BE, we assume it is performed with a ripple-carry adder, which consists of $5n$ gates with a logical depth of $3n$ gates.
In SE, the most performant version comes in the form of scaled addition (e.g., $(a+b)/2$), and is achieved by combining half the bits of bitstream $a$ and half the bits of bitstream $b$.

\section{Evaluation}
\label{sec:evaluation}

To explore the merits and limitations of SCHE, we evaluate the performance, accuracy, and the impact of SIMD parallelization.
In the Palisade library, all 2-input gates have the same performance of roughly $600ms$, and the NOT gate has negligible run time.
Bootstrapping~\cite{palisade-fhew-bootstrapping} is performed after each 2-input gate and accounts for most of the gate's execution time.
Since the performance of all gates are equal, the total compute time required for an application translates directly to the number of gate computations.

\begin{figure}
    \centering
    \includegraphics[width=\columnwidth]{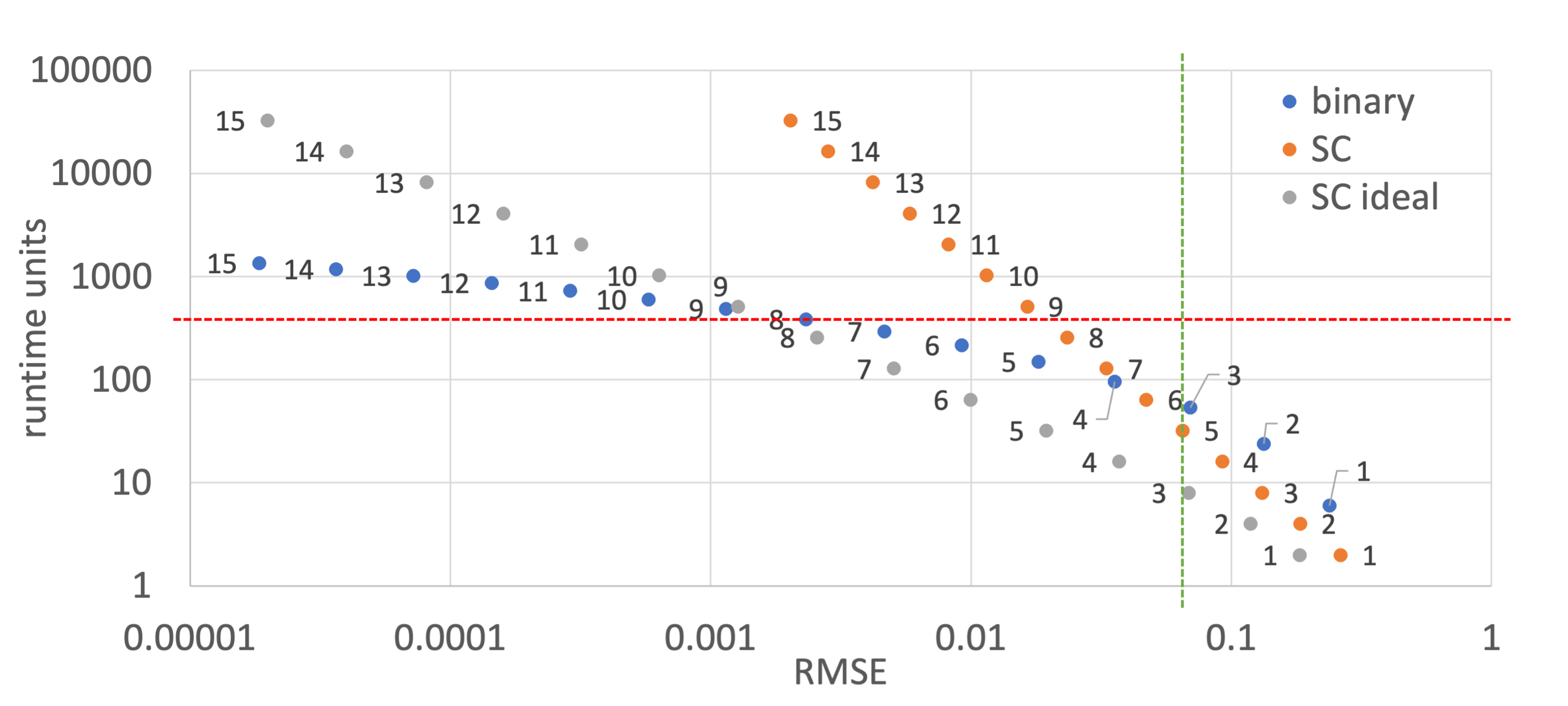}
    \caption{Performance vs. RMSE of array multiplier and SC multiplier at different precision (data labels). 
    }
    \label{fig:mult_perf}
\end{figure}

\noindent\textbf{Total computation time.}
\autoref{fig:mult_perf} compares the performance of SE and BE multiplication assuming no parallelization (i.e., single-threaded).
We plot the total runtime against the root mean square error (RMSE) at each precision, with respect to floating-point multiplication.
The total computation time for $n$-bit precision is on the order of $n^2$ for binary vs. $2^n$ for SC.
If the required computation precision is 8 bits or below (i.e., below the red dashed line in \autoref{fig:mult_perf}), using SE provides performance benefit.
However, because of SC's approximate nature, comparing performance at the same precision may be inadequate if we need to meet a specific accuracy constraint.
As an example indicated by the green dashed line in \autoref{fig:mult_perf}, we may need to consider 5-bits of precision in SE to achieve similar accuracy as 3-bits in BE.
We plot SC ideal in gray to show the best performance-accuracy tradeoff achievable by SC, if we can eliminate correlation and fluctuation errors and the computation is only affected by quantization error.

\begin{figure}
    \centering
    \includegraphics[width=\columnwidth]{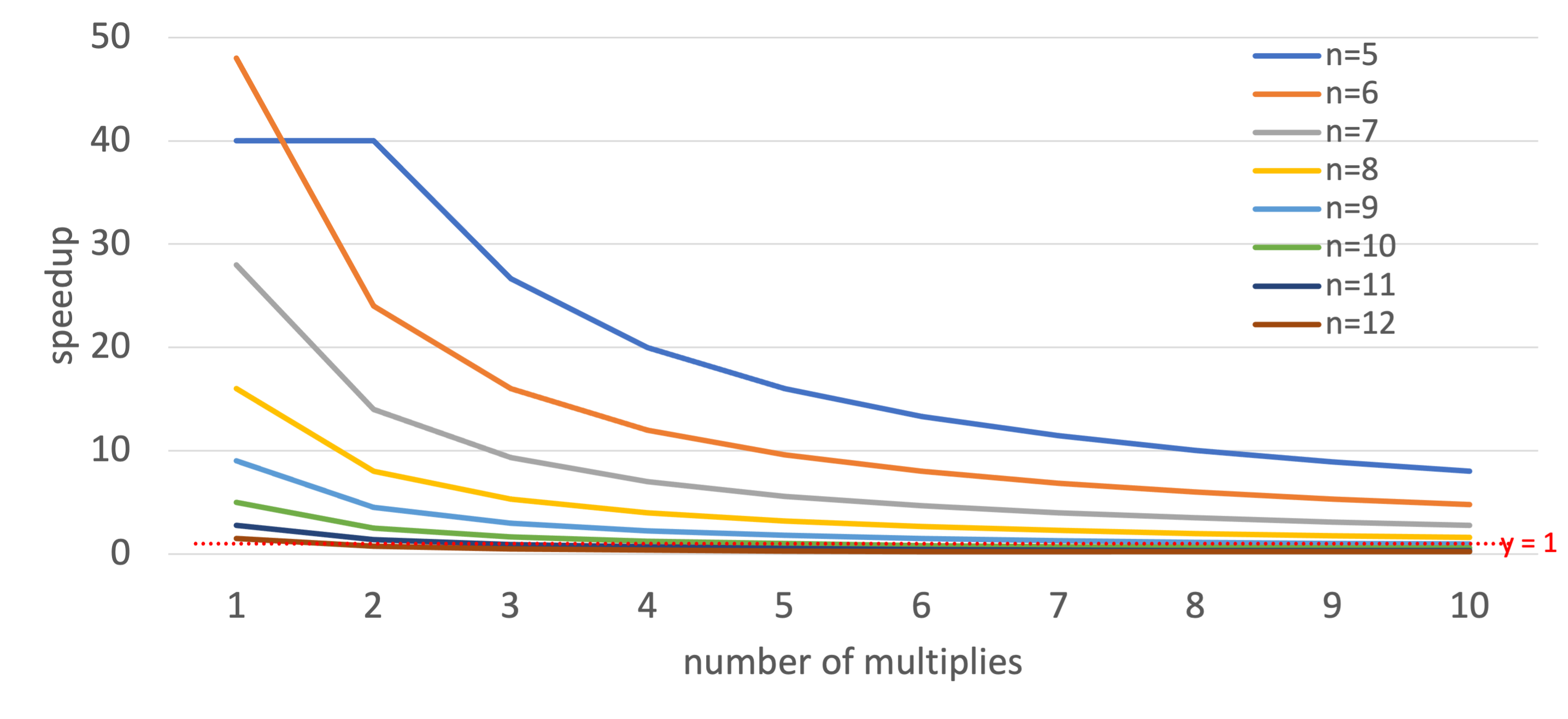}
    \caption{Latency speedup of SC AND multiplier over array multiplier at different precisions. Each line represents a different precision $n$ used.}
    \label{fig:mult_latency}
\end{figure}

\noindent\textbf{Latency.}
If the required computation precision is above 8 bits, SE can still offer performance gains in the form of computation latency, depending on the workload and the underlying hardware's capabilities.
Since an SC multiply is able to be massively parallelized and has shallow circuit depth, if the underlying hardware has enough execution units (e.g., threads, processing elements, etc.), it can finish computation in $600ms$ as opposed to $(n \times 1200)ms$ where $n$ is the input precision.
\autoref{fig:mult_latency} shows the analytical speedup of SCHE over BE multiplication at different precisions and for different numbers of multiplications in the workload using 64 parallel threads (assuming no parallelization overhead). 
The results show that SCHE can provide some gains over BE for special cases where the workload is not large enough for the BE multiply to fully utilize all the execution units.

\begin{figure}
    \centering
    \includegraphics[width=\columnwidth]{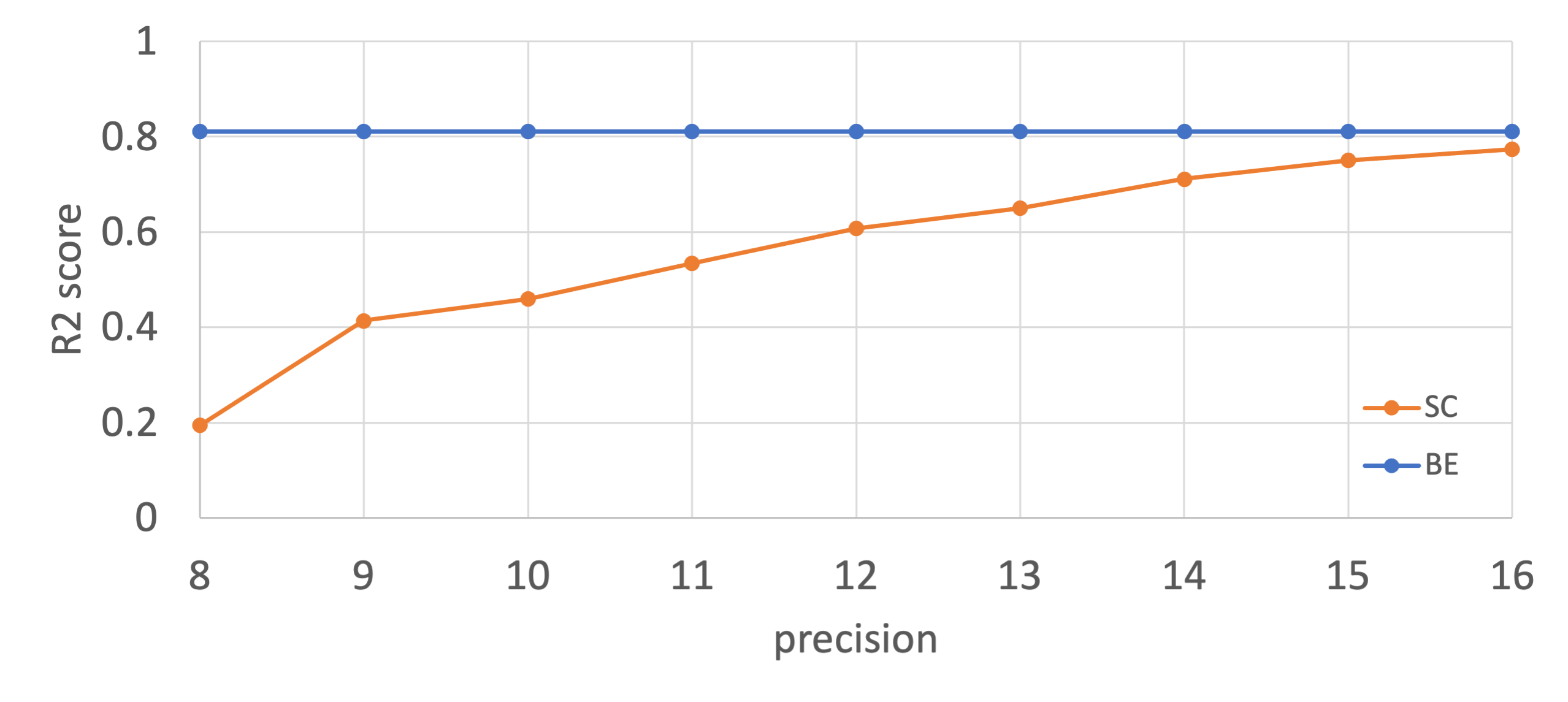}
    \caption{Classification accuracy of SC and BE inference as data precision varies. The baseline floating-point classification accuracy is 0.81.}
    \label{fig:class_accuracy}
\end{figure}

\noindent\textbf{Accuracy tradeoff.}
Since SE provides performance benefits mostly at low precision and incurs some accuracy loss, we evaluate whether the computation accuracy achieved is sufficient for a classification application.
We implement the inference of a classifier that is built with linear regression and thresholding.
The weight and bias of the model is trained using a floating-point representation; the floating-point baseline inference achieves an $R^2$ score of 0.81.
To compare against SC at different precisions, we quantize the model to evaluate the BE accuracy. 
\autoref{fig:class_accuracy} shows the $R^2$ score of the classifier built with SC and compares it to the quantized BE model.
The results show the prediction accuracy is much worse than the BE baseline even at high precision; as a result, we conclude that the performance-accuracy tradeoff of using SE is not worth it. 

\begin{figure}
    \centering
    \includegraphics[width=\columnwidth]{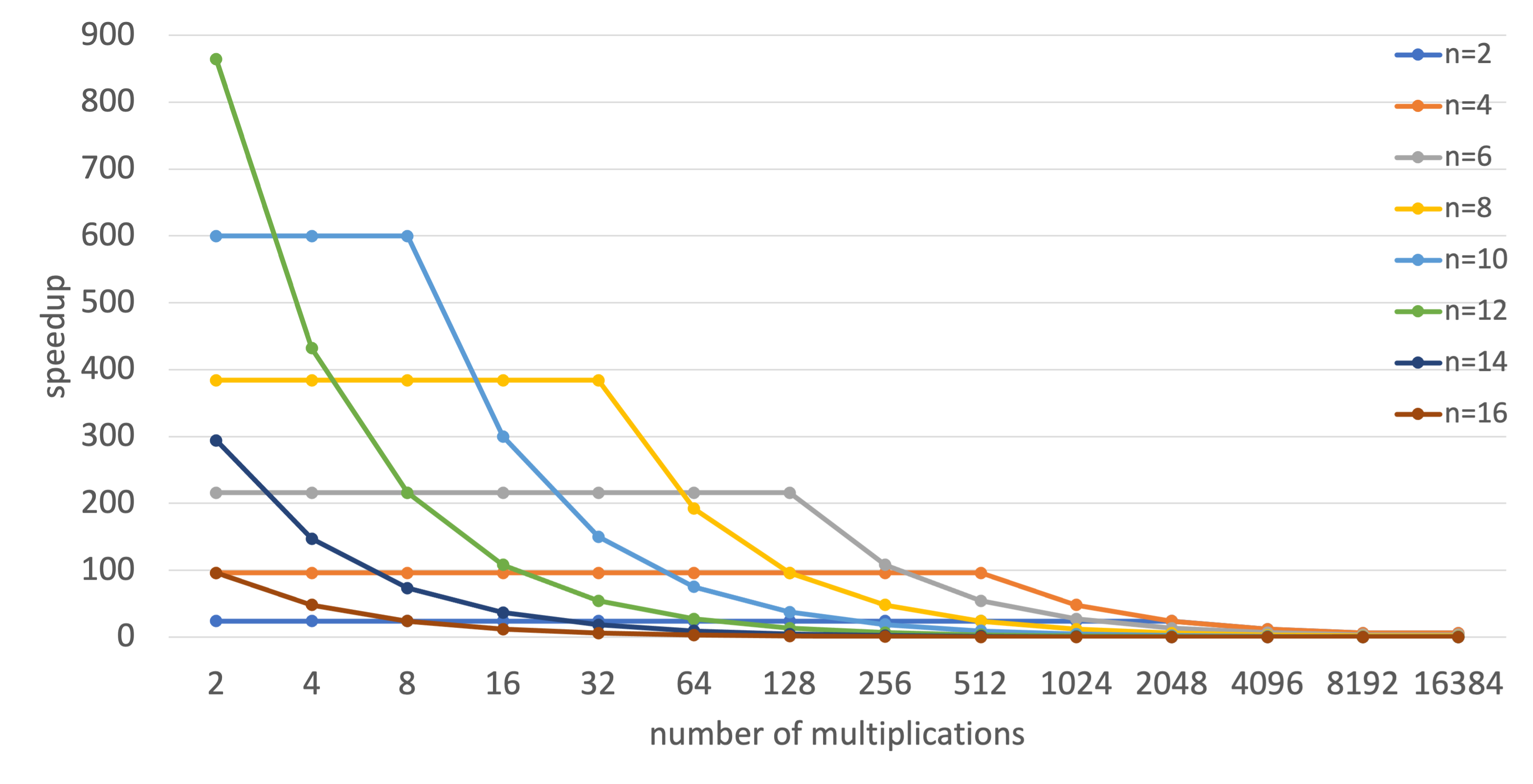}
    \caption{Speedup of SC multiply over BE array multiply for a SIMD vector length of 8192 at different bit precisions $n$. The x-axis represents the number of independent multiplications that an application needs to perform.
    }
    \label{fig:vector_speedup}
\end{figure}

\noindent\textbf{SIMD execution.}
With SE, more opportunities exist for packing and fully utilizing the vector slots of each SIMD execution.
Different bits from a bitstream can trivially be packed into the same vector, and bits from different bitstreams destined for the same operation can be packed into the same vector; this precludes the need for realigning bits in the vectors between operations. 
In contrast, in BE computation bits from the same value cannot easily be packed into the same vector due to data dependencies.
It is possible to predefine a data-layout of which bits can be packed into the same vector based on the schedule of readiness for each gate's input; however, this requires overhead to unpack and realign data for subsequent computation.
To avoid the overhead associated with unpacking and realignment, we can use $n$ vectors of size $m$ to represent $m$ $n$-bit numbers where each vector slot is assigned to one bit of a different number.
However, this scheme's vector utilization is directly tied to the workload size and can significantly impact performance when there are limited number of inputs to process.
\autoref{fig:vector_speedup} shows the speedup of using SC multiply over BE array multiply for a SIMD vector length of 8192.
For a given workload, an increase in SE precision increases the vector slot utilization, and thus the speedup; the speedup is lowered when the precision requires additional vectors to represent its bits.
Because of the high utilization of vector slots, SC execution becomes more feasible at higher precision.

\section{Discussion and Conclusions}
\label{sec:analysis}

After exploring the capabilities of existing HE libraries and evaluating the software performance tradeoffs between SE and BE, we conclude that the strengths and weaknesses of SCHE are very similar to (if not less favorable than) those observed for SC in the hardware context.

First, since we are layering SE on top of HE computation, performance tradeoffs of HE would need to exhibit a different trend than those in CMOS hardware in order to observe different results.
Unfortunately our experiments show that the performance cost of each gate in HE exhibits similar trends as CMOS hardware.
For instance, the performance cost trend for a multiplier in SC compared to BE is the same in CMOS hardware as in HE (i.e., increases exponentially with precision).
As a result, just changing the encoding from BE to SC in HE does not yield a different speedup trend compared to standard CMOS implementations.
If future improvements to the HE libraries create unequal costs depending on the gate type and change this trend, there may be opportunities to re-evaluate the merit of using SCHE.

Second, the fact that computation in HE executes on software libraries as opposed to actual custom hardware that can be spatially parallelized diminishes an important benefit that SC brings.
SC's small arithmetic circuitry allows it to have low area and power, which in turn translates to higher feasibility to spatially parallelized computation.
However, this requires custom hardware implementations to realize all of the benefits which is a non-trivial undertaking.
As a result, by restricting SC to run in software it loses some of the area and power advantages over BE computations and thus loses the spatial parallelism angle.

Third, a few implementation choices in current libraries may limit the effectiveness of SC in HE.
For example, in Palisade's implementation of TFHE, bootstrapping is performed after each gate.
While this allows for computation with high logical depth, it is an unnecessary overhead if the target computation is bounded by shallow logical depth.
We were not able to evaluate the potential of only performing bootstrapping when necessary due to our lack of expertise in the underlying mathematical formulation. 
An expert or compiler that is able to properly select which gates to bootstrap would realize additional performance benefit.
Additionally, Palisade's implementation does not have support for SIMD execution of gates.
Even though its integer HE constructions (e.g., BFV and BGV) support packed encoding (i.e., multiple integers are packed into a vector to allow computation on an element-wise basis between ciphertext vectors), we were unable to emulate Boolean logic with it due to restrictions in setting its plaintext modulus to 2.
Modulo 2 arithmetic would enable an integer encoding of elementary logic gates.
A similar restriction exists in the Microsoft SEAL library~\cite{sealcrypto} where a plaintext modulus of 2 cannot be used with the packed encoding.
If these implementation restrictions are addressed, it would enable other potential SCHE encodings, which may be faster or more efficient.



\balance
\bibliographystyle{ACM-Reference-Format}
\bibliography{references}

\end{document}